# Data on neutron-neutron scattering length extracted in the $n + {}^2H \rightarrow p + n + n$ reaction at 60 MeV

E.S. Konobeevski, A.A. Afonin, A.A. Kasparov, V.V. Mitcuk,

M.V. Mordovskoy, S.I. Potashev, S.V. Zuyev

Institute for Nuclear Research, Russian Academy of Science, Moscow 117312 Russia

E-mail: konobeev@inr.ru

${}^1S_0$ $nn$-scattering length has been measured in the $nd$-breakup reaction at 60 MeV. The experiment was performed in the final state interaction geometry with registration of all three secondary particles. The scattering length $a_{nn} = -16.3 \pm 0.5$ fm was obtained from a comparison of the experimental dependence of the $nd$-breakup yield on the relative energy of $nn$-pair with the simulation results. An analysis of this value of scattering length together with the data obtained in other experiments on $nd$- and $dd$-breakup confirms the hypothesis about the influence of $3N$-forces on the values of $nn$-interaction parameters extracted in these reactions.

## 1. Introduction

One of the important problems of nuclear physics is the principle of charge independence of nuclear forces formulated by V. Heisenberg in 1932. By definition, charge independence is invariance with respect to any rotation in isospin space. Violation of this principle is called charge dependence or charge independence breaking (CIB). Charge symmetry means that in the singlet ${}^1S_0$ state, the proton-proton and neutron-neutron interactions after subtraction of the electromagnetic effects are slightly different. In modern interpretation, the charge dependence of nuclear forces is associated with the mass difference between u- and d-quarks and the electromagnetic interaction between quarks. The most obvious and important case of manifestation of this effect is the neutron-proton mass difference.

The study of the low-energy characteristics of the $NN$ interaction in the singlet spin state – scattering lengths and energies of the virtual ${}^1S_0$ level – plays a special role in determining the measure of the charge symmetry breaking of nuclear forces. Due to the existence of a virtual level with energy ($E_{NN}$) close to zero, the corresponding scattering lengths $a_{nn}$ and $a_{pp}$ are large in magnitude and very sensitive to small differences in the $nn$ and $pp$ potentials.

Accurate experimental data on the scattering lengths and their differences may provide a quantitative estimate of the charge symmetry breaking (CSB) of nuclear forces $\Delta a_{CSB} = a_{pp} - a_{nn}$. Proton-proton scattering length is determined from the *pp*-scattering ($a_{pp} = -17.3 \pm 0.4$ fm) and its accuracy is mainly connected with model-dependent corrections of Coulomb effects. Neutron-neutron scattering length is determined using mostly $n+d \to p+n+n$ and $\pi^- + d \to \gamma + n + n$ reactions investigating the region of the *n-n* final state interaction (FSI) where two neutrons fly together with small relative energy. In [3, 4] it was supposed that significant uncertainty in $a_{nn}$ values obtained in reactions with three particles in the final state [5, 12] may be connected with strong influence of 3*N*-force.

According to authors of Dibaryon Model [13, 14] a new mechanism may arise in this model – scalar meson exchange between the nucleon and dibaryon (singlet). One can propose that such exchange may arise also between two dibaryons (singlets) in the *dd*-breakup reaction. Such additional interaction can lead to a change in the values of $a_{nn}$ and $E_{nn}$ extracted from reactions with two neutrons in the final state. The degree of this change may depend on the relative velocity of the fragments – *nn*-pair and proto*n* or *nn*-pair and diproton for *nd*- and *dd*-breakup, respectively. To estimate this degree we introduce a certain kinematic factor *R*, on which an additional 3*N* interaction may depend [3, 4]. Consider a reaction in which *nn*-pair and a charged fragment fly in the intermediate state. One may calculate the relative velocity of scattering fragments using the kinematics of the two-particle reaction. We choose an arbitrary time interval and determine the distance *R*, to which fragments will fly apart from each other during this time.

One can suppose that the greater is the *R* parameter, the faster the *nn*-pair leaves the region of 3*N*-interaction. So, if the distance *R* between the *nn* pair and proton (or *pp* pair in *dd* reaction) is much larger than the characteristic range of 3*N* interaction, then one can ignore the 3*N*-force contribution in interpretation of the result for $a_{nn}$ value in the given experiment. Let us note that the greatest value *R* = 8.3 fm corresponds to *nd*-breakup experiment at 40 MeV [11].

To test the hypothesis of the dependence of the extracted parameters of the *NN* interaction on the relative distance between the *NN* pair (singlet) and the third particle, additional studies can be performed for few-nucleon reactions for different predicted values of *R*. So one can suppose that studying the *nd*-breakup, for example at 60 MeV (*R* = 10.5 fm), would lead to a smaller influence of 3*N*-force and accordingly to smaller absolute values of *nn* scattering length.

## 2. Statement of the experiment

To confirm these assumptions, we performed a study of the *nd*-breakup reaction at neutron energies above 50 MeV. To study *n-n* final state interaction in this reaction, registration of two neutrons having a very small relative momentum is needed. Neutron–neutron FSI will manifest itself as a peak in the dependence of the reaction yield on the relative energy of two neutrons .

$$\varepsilon = \frac{1}{2}(E_1 + E_2 - 2\sqrt{E_1 E_2} \cos \Theta),  \quad (1)$$

The shape of this dependence $N(\varepsilon)$, according to the Migdal-Watson formula, is sensitive to $E_{nn}$:

$$N(\varepsilon) = A \frac{\sqrt{\varepsilon}}{\varepsilon + E_{nn}}. \quad (2)$$

where $E_{nn}$ is energy of virtual $^1S_0$ state of $nn$-system, which can be obtained by comparing the experimental distribution and simulation results for different values of $E_{nn}$.

In turn, the energy of virtual state $E_{nn}$ is connected with $nn$-scattering length $a_{nn}$ by relation [15]:

$$\frac{1}{a_{nn}} = -\left(\frac{m_n E_{nn}}{\hbar^2}\right)^{1/2} - \frac{1}{2} r_{nn} \frac{m_n E_{nn}}{\hbar^2} + ..., \quad (3)$$

where $r_{nn}$ is the effective radius of $nn$-interaction.

### 3. Kinematical simulation of $n + {}^2H \rightarrow n + n + p$ reaction

To determine the needed experimental conditions and parameters of the setup the kinematical simulation of studied reaction was performed using the programs of simulation of reactions with three particles in the final state [16].

The simulation of the $nd$-breakup yield was carried out in two stages. At the first stage, the formation of pair of neutrons with effective invariant mass $M_{nn} = 2m_n + \varepsilon$ in two-particle reaction $n + {}^2H \rightarrow p + (nn)$ is considered. The dependence of the reaction yield on $\varepsilon$ is taken into account by the number of simulated events with different $\varepsilon$ according to Migdal–Watson formula (2) for a specific $E_{nn}$ value. Thus, the dependence of the form of the reaction yield distribution on $E_{nn}$ value is introduced.

At the second stage, the breakup of the two-neutron-system $(nn) \rightarrow n_1 + n_2$ is considered and the emitting angles and kinetic energies of two neutrons are calculated. From the total number of events being simulated, events that meet the experimental conditions: simultaneous registration of two neutrons emitting at opening angle of 5° in two corresponding neutron hodoscope detectors are selected. The Fig. 1 shows the simulated dependence of the reaction yield on the relative energy for different $E_{nn}$, taking into account all the experimental conditions.

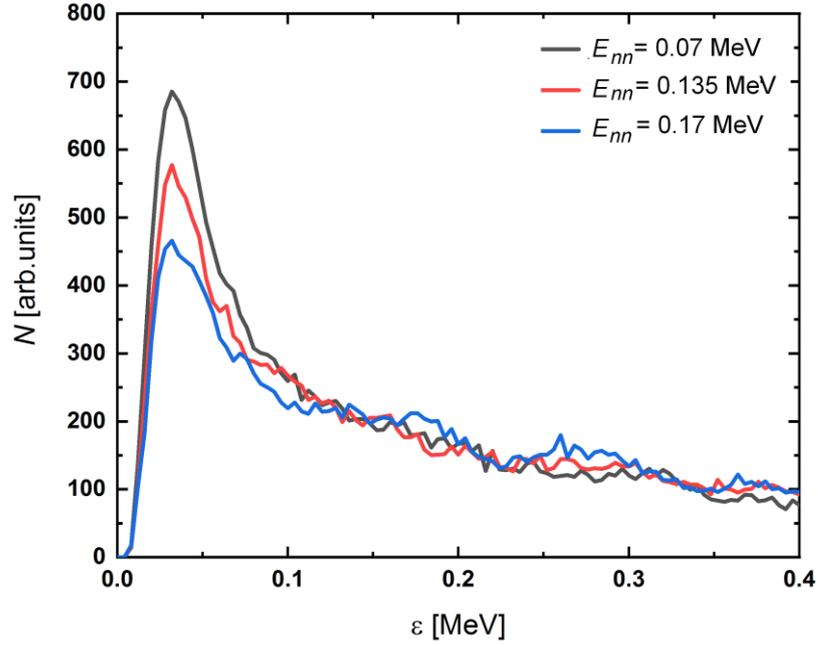

Figure 1: The simulated dependences of the yield of the *nd*-breakup reaction on ε taking into account the experimental conditions: $E_0 = 60$ MeV, $\Theta_{2n} = 60°$, $\Delta\Theta = 5°$ for various values of the virtual state energy $E_{nn}$.

Thus, to obtain data on the energy of the virtual *nn* state in the *nd*-breakup reaction, it is necessary to register the proton and two neutrons and measure the energies of both neutrons at a certain value of their opening angle (for example, $\Delta\Theta \approx 5°$).

## 4. The experimental setup

New data on $E_{nn}$ and $a_{nn}$ values in $n + {}^2H \rightarrow n + n + p$ reaction at 60 MeV were obtained using the neutron beam of the RADEX channel of the Moscow meson factory of the INR RAS. A simplified scheme of the setup is shown in Fig. 2. As a neutron source a beam-stop of 209 MeV protons of the INR linear accelerator was used. The neutrons produced in 60 mm thick tungsten target were collimated at zero angle over 12 m length distance, forming a beam with 50 mm diameter on a measuring deuterium target.

Despite the fact that the spectrum of neutrons incident on a target is wide and includes all energies up to the proton beam energy (~ 200 MeV), simultaneous registration of three particles (proton and two neutrons) in the final state allows to reconstruct the energy of the primary neutron in the $n + {}^2H \rightarrow p + n + n$ reaction for each registered event.

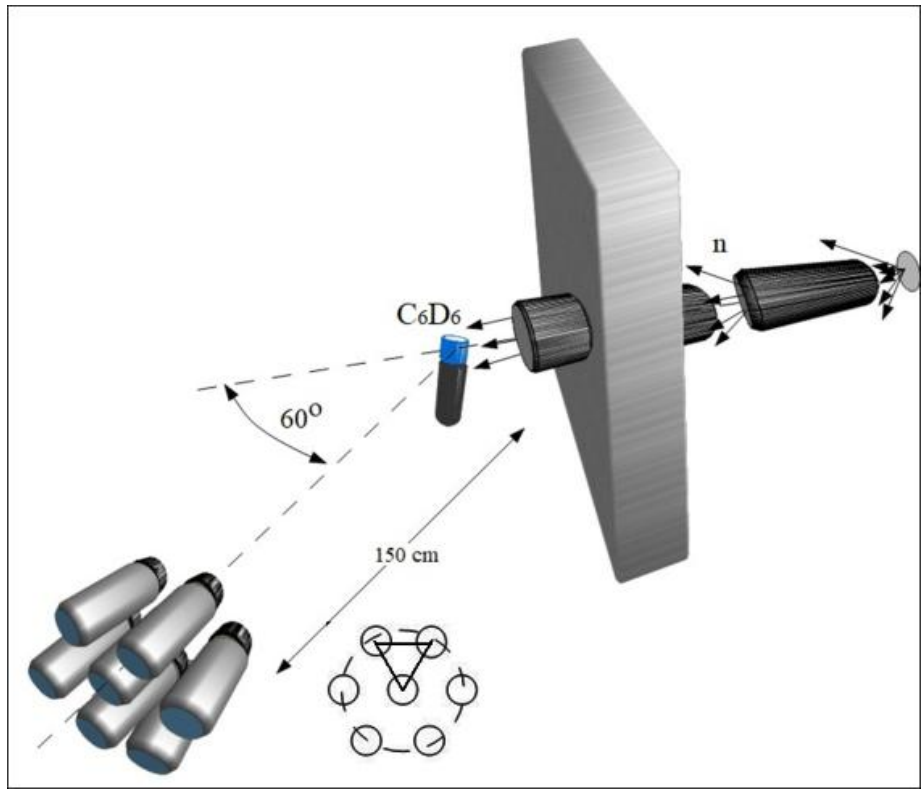

Figure 2: Experimental setup

In the experiment a $C_6D_6$ scintillator (EJ315) was used both as deuterium target and a detector of secondary protons. Secondary neutrons were detected by a hodoscope consisting of seven detectors. The central hodoscope detector was located at $\Theta=60°$ relative to the incident neutron beam axis at 150 cm distance from the deuterium target. The remaining six detectors were located on a circle in the plane perpendicular to the target-central detector direction, while the opening angles between the central and external detectors and between all adjacent external detectors were 5° The neutron energy was determined by the time of flight of neutrons to the detector, while the starting signal of the time-of-flight system was the time signal from an active scintillation target.

In the experiment, signals from the active target (from the secondary proton of the nd-breakup reaction) and from all neutron hodoscope detectors were fed to the inputs of the CAEN-DT5742 digitizer (digital signal processor). A selection of coincident events (the proton and two neutrons in neighboring neutron detectors with $\Delta\Theta=5°$ opening angle) was carried out. For each selected event, the energy of the primary neutron was determined, and thus, the events were sorted by this energy. In particular, this work presents data for the primary neutron energy $E_n = 60 \pm 5$ MeV.

For each registered event, the relative energy of two neutrons was calculated according to (1) from the measured energies of two neutrons and their opening angle. Then the distribution of the number of *nd*-breakup events on ε may be constructed and compared with simulation results.

## 5. Extraction of data on *nn*-scattering length

To determine the energy $E_{nn}$ of the virtual *nn* level (and accordingly scattering length $a_{nn}$), the experimental dependence of the yield of the *nd*-breakup reaction $\frac{dN^{exp}(\Delta\Theta)}{d\varepsilon}$ is compared with the simulation results $\frac{dN^{sim}(\Delta\Theta)}{d\varepsilon}$. In Fig. 3, the experimental data for $\Delta\Theta = 5°$ and incident neutron energy of 60 ± 5 MeV are compared with the simulation results for three energies of the virtual *nn*-state – 0.17 MeV, 0.135 MeV, and 0.07 MeV.

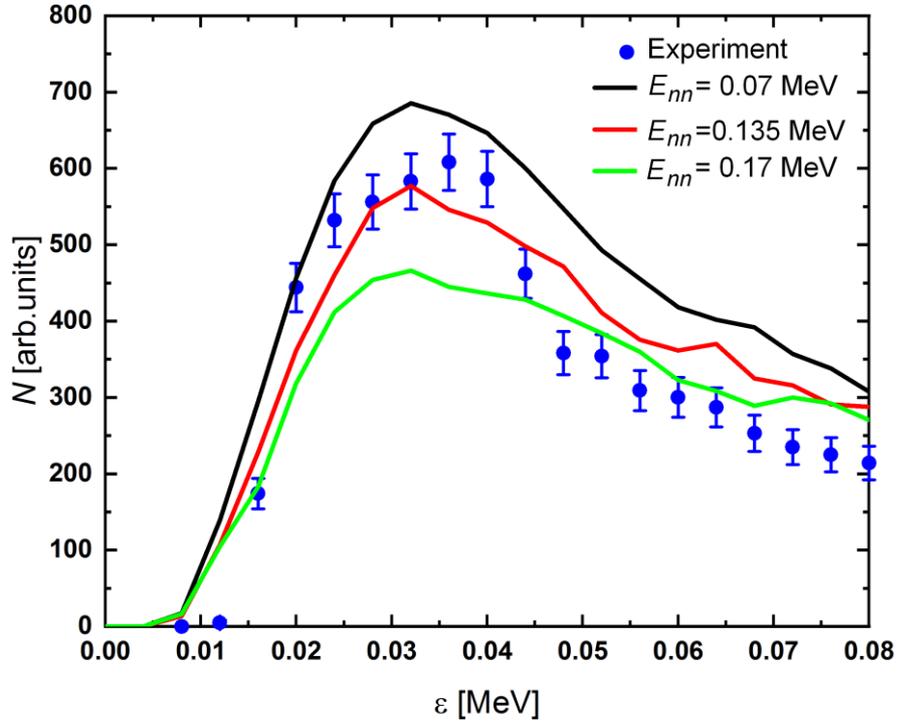

Figure 3: Comparison of the experimental dependence of the reaction yield on ε and the simulation results for various values of $E_{nn}$: 0.07 MeV, 0.135 MeV and 0.17 MeV. The energy of primary neutrons is 60 ± 5 MeV, the opening angle of secondary neutrons $\Delta\Theta = 5°$.

The procedure for determining the energy $E_{nn}$ (and the corresponding scattering length $a_{nn}$) was carried out for the data obtained at the opening angle of 5°. In this case, the simulation was carried out over a wide range of $E_{nn}$ values from 0.07 MeV to 0.24 MeV. Further, the value of $\chi^2$ was minimized for experimental and theoretical (simulated) points:

$$\chi^2(a_{nn}) = \sum_\varepsilon \frac{\left(\dfrac{dN^{exp}(\Delta\Theta)}{d\varepsilon} - A\dfrac{dN^{sim}(\Delta\Theta)}{d\varepsilon}\right)^2}{\left(\Delta\dfrac{dN^{exp}(\Delta\Theta)}{d\varepsilon}\right)^2}, \qquad (4)$$

where $A$ is the normalization coefficient, defined as the ratio of the integrals of the experimental and simulated spectra over a wide range of values of $\varepsilon$ (0 – 0.5 MeV), and $\Delta\dfrac{dN^{exp}(\Delta\Theta)}{d\varepsilon}$ is the statistical error of the experimental points.

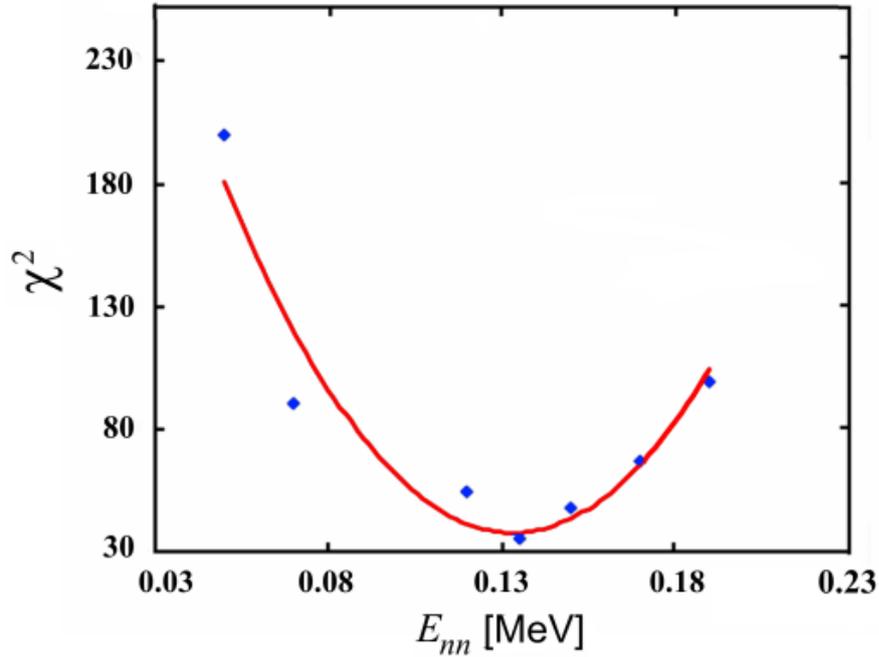

Figure 4: The dependence of $\chi^2$ on $E_{nn}$ for the incident neutron energy $E_n = 60 \pm 5$ MeV and opening angle of secondary neutrons $\Delta\Theta = 5°$. The $\chi^2(E_{nn})$ values were obtained using formula (4) by summing over 20 points over $\varepsilon$ (0.008–0.084 MeV). Solid curve is approximation by a quadratic polynomial.

To find the energy of the virtual level $E_{nn}$ and the statistical uncertainty of its value, the values of $\chi^2(E_{nn})$ are approximated by a quadratic polynomial (Fig. 4). In this case, the minimum value $\chi^2_{min}$ determines the value of $E_{nn}$ and the statistical error in the determination of $E_{nn}$ is given by the formula:

$$\Delta E_{nn} = \left|E_{nn}(\chi^2_{min}) - E_{nn}(\chi^2_{min}+1)\right|. \qquad (5)$$

Thus, for the data presented ($E_n = 60 \pm 5$ MeV, $\Delta\Theta = 5°$) the energy of the virtual neutron-neutron state $E_{nn} = 0.134 \pm 0.007$ MeV was obtained. Although the data were obtained also for other primary neutron energies (for example, for $80 \pm 5$ MeV), however, they were obtained with worse statistics, that did not allow using them to determine the *nn*-scattering length. Using relation (3) with $r_{nn} = 2.83$ fm this value of $E_{nn}$ corresponds to value of neutron-neutron scattering length $a_{nn} = -16.3 \pm 0.5$ fm.

## 6. Analysis of data on *nn*-scattering length

In fig. 5 we can see the data on *nn* scattering length for measurements performed after 1998 [5-12] including our data for *nd*-breakup at 40 MeV [11], *dd*-breakup at 15 MeV MeV [12] and the data obtained in this work at energy of 60 MeV. It should be noted that all the data presented were obtained at different energies and, accordingly, for different values of the *R* parameter.

In [3, 4], it was shown that the experimental data presented can be approximated by a three-parameter exponential function depending on *R*:

$$a_{nn}(R) = a + b \exp(-R/r_0), \qquad (6)$$

whose parameters *a*, *b* and $r_0$ can be obtained from the $\chi^2$ data analysis.

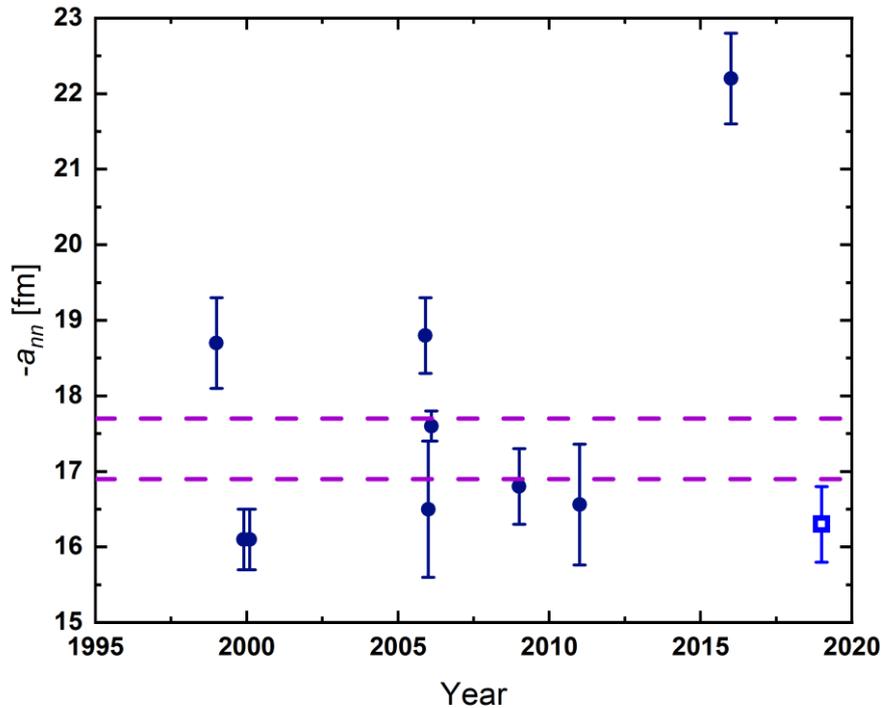

Figure 5: Data on *nn*-scattering length obtained in *nd*- and *dd*-breakup reactions depending on the year of publication. The dashed lines correspond to the limiting values $a_{pp} = -17.3 \pm 0.4$ fm.

Thus, the parameter $a$ determines the asymptotic value of $a_{nn}$ obtained by extrapolating this curve for $R \to \infty$ and therefore being free of the contribution of 3$N$ forces. For the used experimental data in [3, 4], the parameter $a \equiv a_{nn}(\infty) = -15.8 \pm 0.2$ fm was obtained.

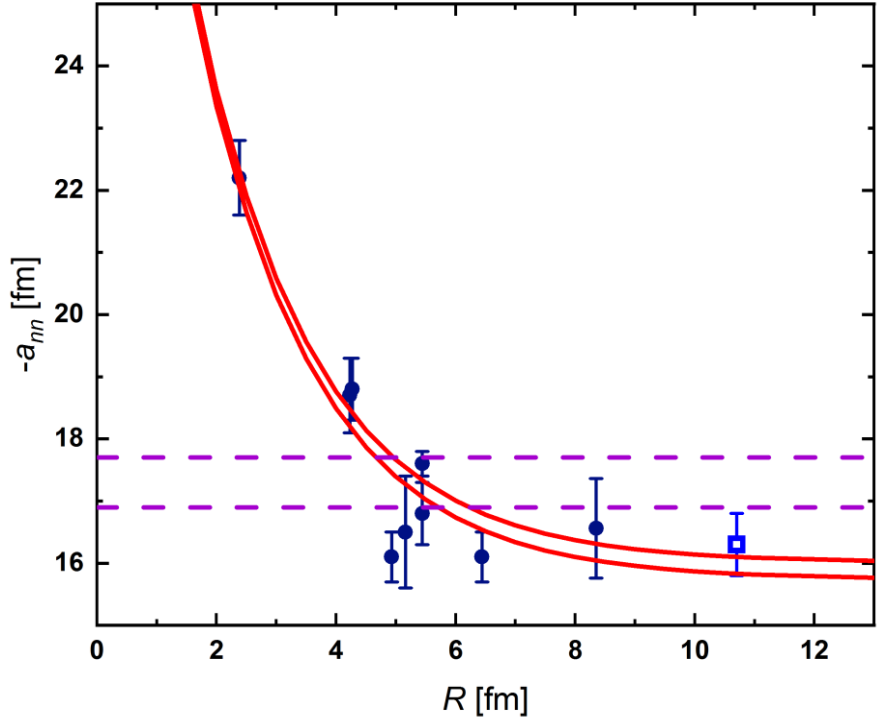

Figure 6: Data on $nn$-scattering length obtained in $nd$- and $dd$-breakup reactions depending on the $R$ parameter. The dashed lines correspond to the limiting values $a_{pp} = -17.3 \pm 0.4$ fm, solid lines – approximation of experimental points by dependence (6) at limiting values $a = -15.9 \pm 0.15$ fm.

Figure 6 presents data on $a_{nn}$ values depending on the values of the $R$ parameter. The value of the neutron-neutron scattering length extracted in the $nd$-breakup reaction at 60 MeV ($R$ = 10.5 fm) is also plotted on the $R$-dependence graph (extreme right point). The $\chi^2$ analysis practically did not change the parameters of the approximating curve $a \equiv a_{nn}(\infty) = -15.9 \pm 0.15$ fm. The obtained experimental value $a_{nn} = -16.3 \pm 0.5$ fm is close to the limiting value (the constant term of the approximation formula), which, in our opinion, indicates that the obtained value of scattering length is close to the value independent on the influence of the 3$N$ interaction.

## 7. Conclusions

Neutron-neutron FSI was studied in the $nd$-breakup reaction in the kinematically complete experiment at neutron energy of 60 MeV. Two neutrons were detected in the kinematic region of FSI at two neutrons opening angle $\Delta\Theta = 5°$. The proton was detected in the active $C_6D_6$ scintillator target. Energies of secondary neutrons were determined by the time of flight, while the proton signal in the active target was used as the starting signal of the TOF system. The primary neutron energy was reconstructed from the kinematics of the reaction, and the relative energy of the *nn* pair was calculated for each event using the values of the neutron energies and their opening angle.

In this experiment the neutron – neutron FSI manifests itself as a maximum in the distribution of the reaction yield on the relative energy of two neutrons, whose shape is sensitive to the value of $E_{nn}$ and accordingly to that of $a_{nn}$. To determine the energy of the virtual *nn* state $E_{nn}$, the experimental dependence of the reaction yield was compared with the simulation results. For the presented data ($E_n$ = 60 ± 5 MeV, $\Delta\Theta$ = 5°), the obtained energy of the virtual neutron-neutron state was $E_{nn}$ = 0.134 ± 0.007 MeV. Using (3) this value of $E_{nn}$ is associated with value of neutron-neutron scattering length $a_{nn}$ = −16.3 ± 0.5 fm. An analysis of this scattering length value in combination with data obtained in other *nd*- and *dd*-breakup experiments confirms the hypothesis about the influence of 3*N* forces on the values of the extracted parameters of *nn*-interaction in these reactions [3, 4].

## References

1. V. G. J. Stoks *et al.,* Phys. Rev. **C49**, 2950 (1994).
2. G. A. Miller, B. M. K. Nefkens, and I. Slaus, Phys. Rep. **194**, 1 (1990).
3. E. S. Konobeevski *et al.,* arXiv:1703.00519v1 [nucl-th].
4. Е. С. Конобеевский *и др.*, ЯФ **81**, 555 (2018) [Phys. Atom. Nucl. **81**, 595 (2018)].
5. D. E. Gonzales Trotter *et al.,* Phys. Rev. Lett. **83**, 3798 (1999).
6. D. E. Gonzales Trotter *et al.,* Phys. Rev. **C73**, 034001 (2006).
7. V. Huhn *et al.,* Phys. Rev. **C63**, 014003 (2000).
8. W. von Witsch, X. Ruan, *and* H. Witala, Phys. Rev. C **74**, 014001 (2006).
9. B. J. Crowe III *et al.,* TUNL Progr. Rep.. **XLV,** 65 (2005-2006).
10. C. R. Howell *et al.,* TUNL Progr. Rep. **XLVIII,** 57 (2008–2009).
11. E.S. Konobeevski *et al.*, Phys. Atom. Nucl. 73, 1302 (2010).
12. E.S. Konobeevski *et al.*, Phys.Atom. Nucl. 78, 643 (2015).
13. A. Faessler et al., J. Phys. G27, 1851 (2001).
14. V. I. Kukulin et al., Ann. of Phys. 325, 1173 (2010).
15. V.A. Babenko, N.M. Petrov, Phys.Atom. Nucl. 76, 684 (2013).
16. S.V. Zuyev, A.A. Kasparov, E.S. Konobeevski, Bull. RAS: Phys. 78, 345 (2014).